%% file: main.tex
\documentclass[12pt]{article}
\usepackage[margin=2.54cm]{geometry}
\usepackage[T1]{fontenc}
\usepackage{lmodern}

\DeclareMathAlphabet{\mathcal}{OMS}{cmsy}{m}{n}

\usepackage{pgfplots}
\usepackage{pgfplotstable}
\pgfplotsset{width=12cm,compat=1.17}

\usepackage{lmodern}
\usepackage{amsmath, amssymb, mathtools}

\usepackage{siunitx}
\usepackage{multirow}
\usepackage{tabularx}
\usepackage{booktabs}

\usepackage{amsmath, amssymb, amsthm, mathrsfs,bm}
\usepackage{booktabs}
\usepackage{url}
\usepackage{graphicx}
\usepackage{enumitem}
\usepackage{algorithm}
\usepackage{algpseudocode}
\usepackage{comment}
\usepackage{xcolor}
\usepackage{lipsum} 
\usepackage{mdframed}  
\usepackage[backend=biber]{biblatex}
\graphicspath{{image}}
\usepackage{setspace}
\usepackage{parskip}
\DeclareFieldFormat{title}{#1}

\usepackage{titling}
\usepackage{fancyhdr}
\newcommand{\articletype}[1]{\textit{\small #1}\par\vspace{2mm}}

\fancypagestyle{firstpage}{%
  \fancyhead[C]{\includegraphics[height=3.5cm, width=4cm]{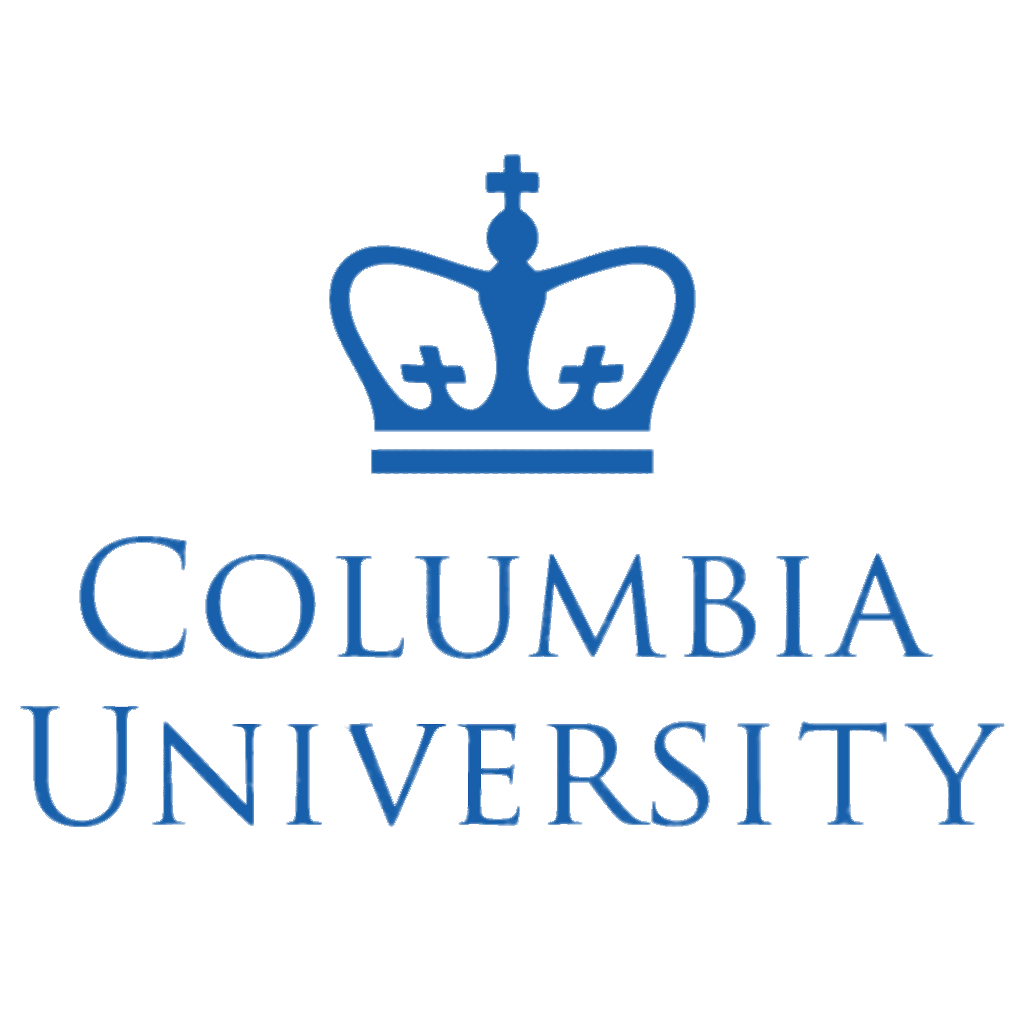}}  

}

\renewenvironment{abstract}
  {\begin{mdframed}[
    backgroundcolor=gray!20, 
    innertopmargin=10pt, 
    innerbottommargin=15pt, 
    innerleftmargin=15pt, 
    innerrightmargin=15pt, 
    linewidth=0pt, 
    roundcorner=0pt,
    frametitleaboveskip=5mm, 
    frametitlebelowskip=-1mm, 
    frametitle={\bfseries Abstract}, 
    frametitlefont={\bfseries},
    frametitlealignment=\raggedright
    ]
   \vspace{2mm}}
  {\end{mdframed}}

\newcommand{\keywords}[1]{
  \vspace{12pt} 
  \par\noindent\textbf{Keywords:} #1
}

\title{\articletype{} A multi-factor market-neutral investment strategy for New York Stock Exchange equities}

\author{G. M. Gkolemis \hspace{25pt} A. R. Lee \hspace{25pt} A. Roudani\\
{\footnotesize Contact: g.gkolemis@columbia.edu / arl2244@columbia.edu / amine.roudani@columbia.edu}\\
{\footnotesize Author affiliation: Columbia University, New York, NY, USA}
}

\date{December 2024}

\begin{document}

\maketitle
\thispagestyle{firstpage}

\begin{abstract}
This report presents a systematic market-neutral, multi-factor investment strategy for New York Stock Exchange equities with the objective of delivering steady returns while minimizing correlation with the market. A robust feature set is integrated combining  momentum-based indicators, fundamental factors, and analyst recommendations. Using various statistical tests for feature selection, the strategy identifies key drivers of equity performance and ranks stocks to build a balanced portfolio of long and short positions. Portfolio construction methods, including equally weighted, risk parity, and minimum variance beta-neutral approaches, were evaluated through rigorous backtesting. Risk parity demonstrated superior performance with a higher Sharpe ratio, lower beta, and smaller maximum drawdown compared to the S\&P 500. Risk parity's market neutrality, combined with its ability to maintain steady returns and mitigate large drawdowns, makes it a suitable approach for managing significant capital in equity markets.
\end{abstract}


\input{Introduction}

\input{Data}
\input{Results}
\input{Discussion}
\input{Summary}

\section*{Acknowledgements}
We would like to thank Naftali Cohen, Columbia faculty member and quant on Wall Street, for his invaluable advice.

\section*{Reproducibility}
For readers interested in reproducing or experimenting with this strategy, the following link contains the Python code that was used: https://github.com/amineroudani/Data-Driven-Methods-in-Finance-Final-Project.

\printbibliography
\end{document}

%% file: Introduction.tex
\section{Introduction}
Multi-factor models have been very popular both in the professional world of investing as well as in research \cite{arshanapalli1998multifactor, blitz2019characteristics, fama1996multifactor}. The consistent performance of such strategies and their interpretability makes them attractive for investors. This paper investigates such a model and demonstrates the methodology of building a multi-factor strategy, from feature engineering and selection to portfolio construction and backtesting.

On a high level, a multi-factor model calculates expected stock returns based on a number of factors that have explanatory power with regard to stock returns. In order to identify the most significant factors to include in the model, a number of statistical tests ought to be performed. These tests shall determine which factors can best predict future returns, while ensuring that all chosen factors are uncorrelated with each other. Having identified the most significant features, the expected returns of the stocks in our investment universe can be computed. Based on these returns a portfolio can be constructed that will allocate the capital at the investor's disposal between a number of selected stocks.

This process is explained in detail in the following sections of this paper. Section 2 presents the sources and range of our data as well feature selection and portfolio construction. The performance of the resulting model is shown in Section 3, where both in-sample and out-of-sample results are analyzed. Section 4 discusses interesting aspects of the strategy's performance as well as limitations and suggestions for future improvement. Finally, Section 5 contains some concluding remarks.

%% file: Data.tex
\section{Methodology}

\subsection{Data sources and range}
All data used in the design and testing of the quantitative strategy described in this article were extracted from the Wharton Research Data Services (WRDS) databases. There are four sets of data that were utilized: NYSE daily equity prices, company monthly fundamental factors, analyst recommendations at various frequencies and S\&P 500 index daily prices. This information comes from three different vendors that make their data available on the WRDS platform: Compustat, CRSP and IBES. The specific tables accessed by SQL queries are described in Table \ref{tab:wrds}.

Information from different data frames was combined using each stock's CUSIP (Committee on Uniform Securities Identification Procedures) number. CUSIPs are 9-character identifiers that capture an issue's important differentiating characteristics within a common structure \cite{cusip}; in other words, they are unique stock identifiers. This paper focuses exclusively on NYSE stocks due to the availability of information and diversity in market cap and industries the NYSE offers.

\begin{table}[h]
\centering
\begin{tabular}{lll}
\textbf{Data vendor} & \textbf{Table name on WRDS} & \textbf{Table contents} \\ \hline
Compustat & $comp\_na\_daily\_all.secd$ & Equity prices  \\ 
CRSP/ Compustat & $wrdsapps\_finratio.firm\_ratio$ & Fundamental factors\\ 
IBES & $tr\_ibes.recddet$ & Analyst recommendations \\ 
Compustat & $comp\_na\_daily\_all.idx\_daily$ & S\&P 500 index \\ \hline
\end{tabular}
\caption{Location of data used on the WRDS platform.}
\label{tab:wrds}
\end{table}

We separated equity data into three sets: training, validation, and test. The training and validation sets are in-sample data used to build, experiment with, and refine the strategy presented in this paper. The test set was used at the end of the research to confirm that our model also performs robustly on out-of-sample data. In particular, the training set ranges from 2000 to 2010, validation took place from 2011 to 2015 and the testing was performed on the years between 2016 and 2024.

\subsection{Feature Engineering}
Precautions were taken to avoid leakage between in-sample and out-of-sample data. Furthermore, within each sample, we ensured that the data remained Point-In-Time (PIT). When a dataset is characterized as PIT, each data point only contains information that would have been available at the respective date, ensuring reproducibility in real life and robustness.

\subsubsection*{Momentum-based Features}
Three momentum-based features features were computed from the equity price data: Relative Strength Index (RSI), True Strength Index (TSI) and trended momentum.

RSI is a term often used to highlight the relative strength of a security in relation to the market on which it is traded or with a different security \cite{rsi}. It evaluates overbought (higher RSI) or oversold (lower RSI) conditions of an asset. It is calculated using the following formula:
    \[
    RSI = 100 - \frac{100}{1 + RS}
    \]
Where:
\[
RS = \frac{\text{Average Gain over } n \text{ periods}}{\text{Average Loss over } n \text{ periods}}
\]

TSI is a technical momentum oscillator that helps traders identify the strength and direction of price movements \cite{tsi}. Similar to RSI, it is used to identify overbought/ oversold assets and trend reversals. The True Strength Index (TSI) is calculated as:
\[
TSI = 100 \times \frac{ EMA(\Delta P)}{EMA(EMA(|\Delta P|))}
\]
Where:
\[
\Delta P = P_t - P_{t-1}
\]
- \( \Delta P \): Price Change.\\
- \( EMA(\cdot) \): A function that returns its argument with exponential moving average applied. 

The third feature that was engineered is trended momentum and has the following rationale. In general, momentum is defined as the change in price between the beginning and the end of some time period. However, this definition overlooks the price movements between those two points. As shown in Figure ~\ref{fig:stock_prices}, two stocks with the same momentum can have completely different price charts. In other words, the common definition of momentum does not tell the entire story. The idea is that "if investors engage in trend-chasing, a clear trend would induce more of such behavior due to the reduced cognitive load required to process that information" \cite{trendclarity}. To capture this idea, we take the daily price time series over the past 12 months (excluding the previous month to avoid reversion) and run a simple linear regression

\[
y_{i,t} = \beta_i x_t + \alpha_i + \epsilon_{i,t}
\]

where:

- \( y_{i,t} \) is the price of stock \( i \) at time \( t \), \\
- \( x_t \) is the time index \( t = 1, 2, \ldots, T \),\\
- \( \beta_i \) is the slope of the trend line for stock\( i \), \\
- \( \alpha_i \) is the intercept, \\
- \( \epsilon_{i,t} \) is the error term

The slope \( \beta_i \) is estimated using the least squares method as:

\[
\hat{\beta}_i = \frac{\sum_{t=1}^T (x_t - \bar{x})(y_{i,t} - \bar{y}_i)}{\sum_{t=1}^T (x_t - \bar{x})^2}
\]

where:

- \( \bar{x} \) is the mean of the time indices: \( \bar{x} = \frac{1}{T} \sum_{t=1}^T x_t \),\\
- \( \bar{y}_i \) is the mean of the stock prices: \( \bar{y}_i = \frac{1}{T} \sum_{t=1}^T y_{i,t} \).

The goodness of fit of the regression is captured by the coefficient of determination \( R^2 \), which is given by:

\[
R^2_i = 1 - \frac{\sum_{t=1}^T (y_{i,t} - \hat{y}_{i,t})^2}{\sum_{t=1}^T (y_{i,t} - \bar{y}_i)^2}
\]

where:

- \( \hat{y}_{i,t} = \hat{\beta}_i x_t + \hat{\alpha}_i \) is the predicted price,\\
- \( \sum_{t=1}^T (y_{i,t} - \hat{y}_{i,t})^2 \) is the sum of squared errors (SSE),\\
- \( \sum_{t=1}^T (y_{i,t} - \bar{y}_i)^2 \) is the total sum of squares (TSS).

The slope \( \hat{\beta}_i \) measures the strength and direction of the trend, while \( R^2_i \) indicates how well clear the trend is to the human eye. To combine these two metrics into a single measure of trend clarity, we define the Trended Momentum \( TM_i \) as:

\[
TM_i = \hat{\beta}_i \times R^2_i
\]


\pgfplotstableread[col sep=comma]{stock_time_series.csv}\stockdata

\begin{figure}[htbp]
    \centering
    \begin{tikzpicture}
        \begin{axis}[
            title={Stock Price Time Series},
            xlabel={Time (Days)},
            ylabel={Price},
            width=0.6\textwidth,
            height=0.4\textwidth,
            line width=1pt,
            tick style={line width=0.6pt},
            xmin=0,
            xmax=200,
            ymin=90,
            ymax=160,
            xtick={0,50,100,150,200},
            ytick={100,120,140,160},
            xticklabel style={rotate=0},
            yticklabel style={/pgf/number format/fixed},
            axis lines=box,
            legend style={at={(0.5,-0.25)}, anchor=north, legend columns=-1, draw=none},
            legend cell align={left}
        ]

        \addplot[color=blue, thick] table[x=Time, y=Stock_A] {\stockdata};
        \addlegendentry{Stock A}

        \addplot[color=red, thick] table[x=Time, y=Stock_B] {\stockdata};
        \addlegendentry{Stock B }

    \end{axis}
    \end{tikzpicture}
    \caption{Comparison of Stock A (clear trend) and Stock B (less clear trend), both with the same 'momentum'}
    \label{fig:stock_prices}
\end{figure}
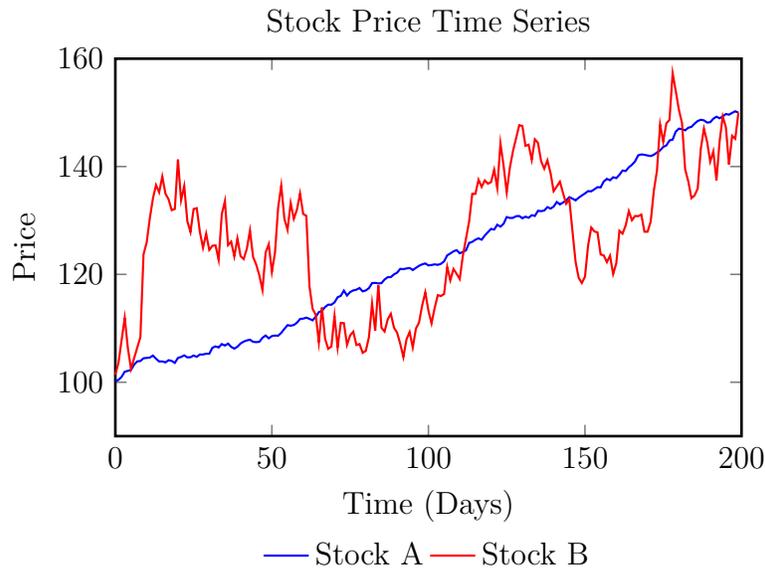

\subsubsection*{Fundamental factors}
With the goal of measuring behavioral biases, simple feature engineering was applied to the fundamental factors. We computed the percentage change between the most recent value, and its 6 month rolling average. Using a rolling average reduces noise and using percentage change, as opposed to absolute change, makes the feature comparable cross-sectionally between companies.

\subsubsection*{Analyst recommendations}
Analyst recommendations received the same treatment as fundamental factors. The change with respect to the previous observation for each stock was calculated. Again, the computation was performed in a way that ensured our data remained PIT. Recommendations were integrated in the dataset only after their announcement day.

\subsection{Feature Selection}
Extracting equity data and engineering various features, as discussed above, created a total of 162 factors. In order to make the value of these features comparable to each other, the z-scores of each feature were calculated (data was grouped by date to avoid violating PIT principles) and capped at minimum/ maximum values of -3/ 3 to ensure our model is robust to outliers. A series of quantitative and qualitative tests were, then, run to identify which factors were most significant and predictive of future returns:
\begin{enumerate}
    \item The correlation of each factor with monthly returns was calculated for an initial filtering. Features with absolute correlation values lower than 0.01 were discarded.
    \item Individual regressions between each remaining metric and monthly returns were run. Factors whose coefficient sign was contrary to conventional economic theory were also rejected. For example, gross profit margin exhibited a negative coefficient, even though, in theory, companies with high margins should expect superior returns. \item After ensuring that the remaining features are not correlated to each other, a multivariate regression between the factors and monthly returns is implemented. Metrics with high coefficients and low p-values are preserved and are the basis of the multi-factor model. 
\end{enumerate}

Out of an initial 162 factors, 11 are found to have predictive power in relation to monthly stock returns. The resulting feature selection is the following: RSI, R\&D expenses divided by Sales (R\&D/S), Sales-to-Price (S/P), Book-to-Market (B/M), the monthly change of B/M (B/M change), Return on Assets (ROA), Accrual divided by Assets (A/A), Sales-to-Equity (S/E), monthly change of S/E (S/E change), monthly change of asset turnover (AT-TV change), momentum.

\subsection{Stock ranking}
Having identified the most significant factors, a model is set up to calculate expected stock returns. Our strategy executes trades on a monthly basis and, therefore, the objective is to forecast which equities will perform best and worst over the coming month. At the beginning of each month, a Lasso regression between monthly returns and the 11 selected factors is run over the past year to identify the feature coefficients. This monthly calculation of the weights reflects the reality of constantly evolving market conditions. After computing the feature weights, expected returns for all NYSE stocks over the next month are calculated and ranked. The top 40 stocks are selected for long positions and the bottom 40 for short positions.

Choosing Lasso regressions and picking the top and bottom 40 stocks are two decision worth elaborating on. As mentioned above, market conditions continuously change. Therefore, while the factor selection was robustly implemented, a regression model that has an in-built feature selection aspect was an attractive choice. Regarding the selection of 40 stocks, this number was considered optimal, as it provides sufficient diversification while ensuring that the impact of the best-performing stocks is not overly diluted.

\subsection{Portfolio construction}
The selected stocks are used to construct a monthly rebalanced portfolio. Here we demonstrate three different approaches: equally weighted, risk parity and minimum variance beta neutral.

\subsubsection*{Equally weighted portfolio (EWP)}
An EWP simply allocates an equal amount of money to every stock. This means that every selected stock is allocated a weight of 1/80, since there are 40 long positions and 40 short positions. 

\subsubsection*{Risk parity portfolio (RPP)}
A RPP allocates weights in a manner such that each asset contributes equally to the total risk of the portfolio. The rationale behind this portfolio construction method is that, instead of assigning equal dollar amounts to each position, equal amount of risk exposure is assigned to each position. In our case, RPP is implemented in the context of convex optimization with the help of Python's cvxpy library.

The optimization problem for the RPP \cite{jeancharles2013rpp} can be formulated as:
\[
\min_{x} \left( \frac{1}{2} x^\top \Sigma x - \sum_{i=1}^n \frac{1}{n} I \log(x_i) \right)
\]
Subject to:
\[
x_i \geq 0 \quad \text{for all } i = 1, 2, \dots, n.
\]
Where:
\begin{itemize}
    \item \( x \): Portfolio weight vector to be optimized.
    \item \( \Sigma \): Covariance matrix of asset returns.
    \item \( I\): A unit vector.
    \item \( \log(x_i) \): Logarithmic barrier term ensuring weights align with \( c \).
\end{itemize}
The solution \( x^* \) is normalized to compute the portfolio weights:
\[
w_i = \frac{x_i^*}{\sum_{j=1}^n x_j^*}.
\]

\subsubsection*{Minimum variance beta-neutral portfolio (BNP)}
The focus of this process is on constructing a beta-neutral portfolio, a portfolio that will not be correlated with the movement of the market. Again, a convex optimization problem is formulated with the help of Python's cvxpy, where the variance of the portfolio is minimized under the constrained that portfolio beta should be zero. The betas of individual stocks are computed by implementing regressions between the stock price and the S\&P 500 over three years.

The optimization problem for the market-neutral minimum variance portfolio is formulated as:
\[
\min_x \frac{1}{2} x^\top \Sigma x
\]
Subject to:
\[
x_i \geq 0, \quad \text{for all } i = 1, 2, \dots, n.
\]
\[
\beta^\top x = 0
\]
\[
\sum_{i=1}^n x_i = 1
\]
Where:
\begin{itemize}
    \item \( x \): Portfolio weight vector (\( n \times 1 \)).
    \item \( \Sigma \): Covariance matrix of returns (\( n \times n \)).
    \item \( \beta \): Vector of asset betas relative to the market.
\end{itemize}

The solution \( x^* \) represents the optimal portfolio weights, where long positions are non-negative and short positions are adjusted to have negative weights.

\subsection{Backtesting}

\begin{figure}[h!]
\centering
\includegraphics[width=1\textwidth]{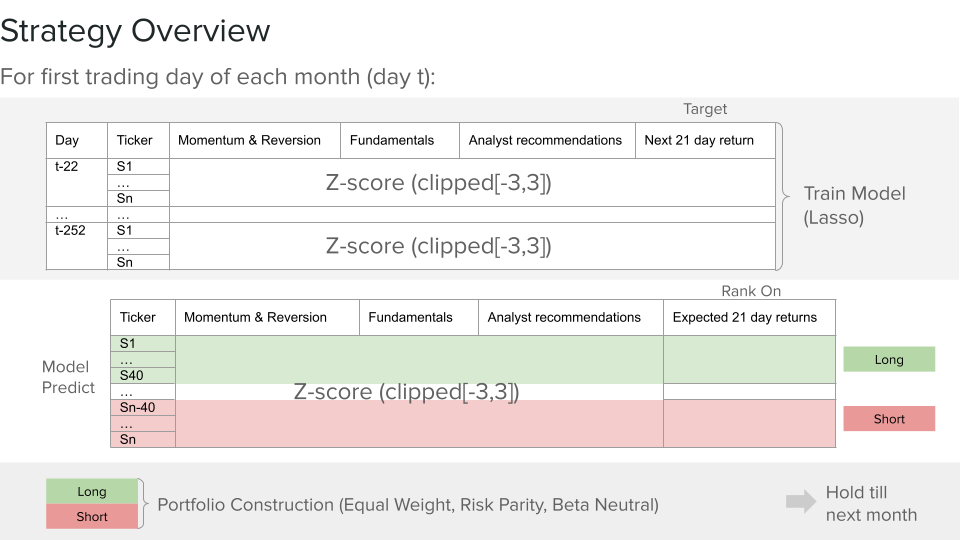}
\caption{Strategy Overview}
\label{fig:Backtesting}
\end{figure}

The flow of the entire investment strategy can be summarized in Figure \ref{fig:Backtesting}. The backtesting of the strategy was conducted on the basis of this flow:
\begin{itemize}

\item  On first trading day of each month, historical features of past 22 to 252 days are used to fit a Lasso-Regression model against their respective future 21 day returns. Note that only features before the past 21 days are used. Otherwise, the target future 21 day return will contain today's return, which would cause information leak.

\item Prior to fitting the model, features are grouped by days to compute z-scores, which are clipped between [-3,3] to confine extreme values.

\item Then, the trained model will provide expected future 21 days return of each stocks based on features observed today. Stocks are ranked on this expected return. The top 40 stocks are chosen to enter a long position, and the bottom 40 are chosen to be shorted.

\item Finally, a portfolio (EWP, RPP, or BNP) will be constructed with the 80 chosen stocks. This position will be held for a month before we repeat this same process.

\end{itemize}

%% file: Results.tex
\section{Results}
This section is organized as follows: first the training and validation sets are presented and based on these results a specific portfolio construction method is chosen for the test set and for our strategy in general.

\begin{figure}[h!]
\centering
\begin{minipage}{0.49\textwidth}
    \centering
    \includegraphics[width=\textwidth]{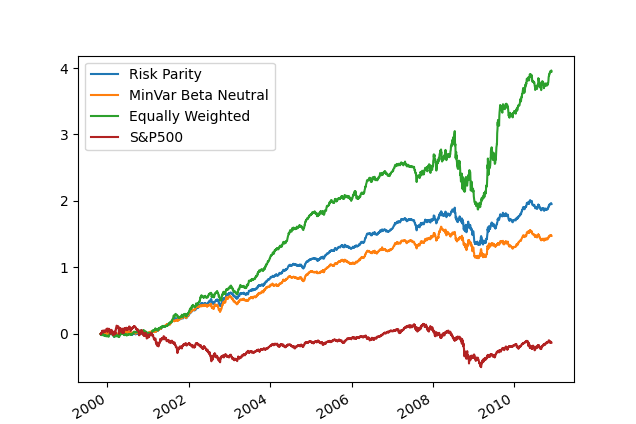}
    \caption{Training set}
    \label{fig:trainingreturns}
\end{minipage}
\hfill
\begin{minipage}{0.49\textwidth}
    \centering
    \includegraphics[width=\textwidth]{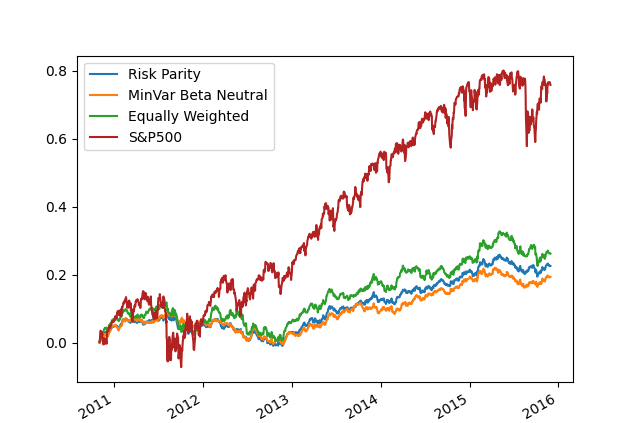}
    \caption{Validation set}
    \label{fig:validationreturns}
\end{minipage}
\end{figure}

Figures \ref{fig:trainingreturns} and \ref{fig:validationreturns} demonstrate the returns obtained by the three portfolios and the S\&P 500. We observe that the risk parity and minimum variance beta-neutral portfolios secure stable returns without excessive movements either upward or downward. The equally weighted portfolio shows high variance in the period of the 2008 financial crisis in the training set but follows a similar path as the other two portfolios during validation. Regarding the S\&P 500, it performs poorly with negative cumulative returns over the whole training set, but outperforms the three portfolios in nominal returns.

Table \ref{tab:validation} lists each strategy's Sharpe ratio, beta and maximum drawdown on the validation set. Risk parity and minimum variance exhibit a higher Sharpe than the market benchmark in combination with values of beta close to zero. Even though the beta of the risk parity and equally weighted portfolio is not explicitly programmed to approach zero, both portfolios are observed to be market neutral.

Since risk parity achieved the highest Sharpe ratio, lowest absolute value of beta and a small maximum drawdown, it was selected as the preferred portfolio construction method for our strategy.

\begin{table}[h!]
\centering
\begin{tabular}{lcccc}
 & \textbf{Risk Parity} & \textbf{Min Var} & \textbf{Equally Weighted} & \textbf{S\&P 500} \\ \hline
Sharpe     & 0.89 & 0.84  & 0.76 & 0.80 \\ 
Beta & -0.002  & 0.063  & -0.007 & 1 \\ 
Max drawdown & -8.82\%  & -7.79\% & -10.22\%  & -19.39\% \\ 
\end{tabular}
\caption{Validation set results (in-sample)}
\label{tab:validation}
\end{table}

\begin{figure}[h!]
\centering
\includegraphics[width=0.8\textwidth]{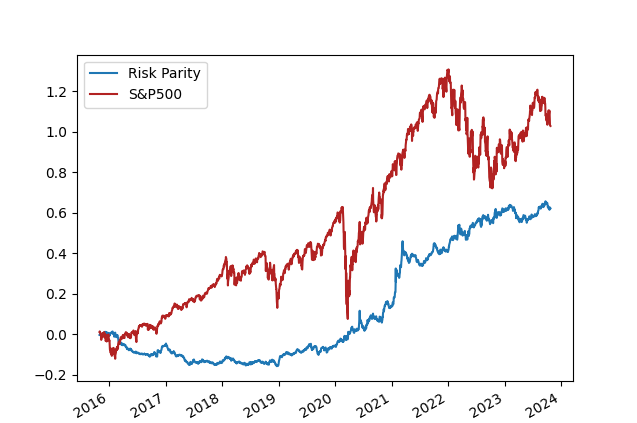}
\caption{Out-of-sample results: risk parity is the chosen portfolio construction method.}
\label{fig:test}
\end{figure}

The out-of-sample results are obtained by implementing the strategy between 2016 and 2024, illustrated in Figure \ref{fig:test}. The figure shows that the risk parity portfolio achieves small and slightly negative returns at the beginning before starting on an upward path. Again, the stability of risk parity is observed; during the COVID crash and other periods of sharp market declines (or hikes), our strategy remains unaffected and stays on an almost linear path. This is a highly desired behavior trait especially for portfolios that manage substantial sums of capital. An ideal such portfolio has low variance and steady returns.

The details of the metrics achieved can be seen in Table \ref{tab:test}. The strategy's Sharpe ration is higher than that of the S\&P 500 for the same period and returns are not correlated with the market, rendering the strategy market neutral. The maximum drawdown is smaller than the market benchmark.

\begin{table}[h!]
\centering
\begin{tabular}{lcc}
 & \textbf{Risk Parity} & \textbf{S\&P 500} \\ \hline
Sharpe     & 0.81 & 0.57 \\ 
Beta       & 0.007  & 1 \\ 
Max drawdown & -16.87\%  & -33.93\% \\ 
\hline
\end{tabular}
\caption{Out-of-sample metrics for Risk Parity and S\&P 500.}
\label{tab:test}
\end{table}

%% file: Discussion.tex
\section{Discussion}

\subsection{Performance}

The market neutral strategy proposed displayed a Sharpe ratio of 0.89 and 0.81 in the validation and  test period respective. This is higher than the Sharpe ratio displayed by the S\&P 500 in the same periods, which are 0.80 and 0.57. In terms of maximum drawdown, or strategy experienced a 8.82\% and 16.87\% maximum drawdown in the validation and test period. The maximum drawdown is smaller compared to the 19.39\% and 33.93\% experience by the S\&P 500. Finally, in both periods, our strategy displayed a small beta of -0.002 and 0.007.

Overall, our market neutral strategy appears to be less volatile and less risky than the S\&P 500. It can also maintain a higher Sharpe ratio in caomparison to the S\&P 500.

An interesting period within the test period is 2016 - 2019, when our strategy experienced the 16.87\% maximum drawdown. In this period, the return of our strategy is not too volatile. However, it experienced a rather steady loss over the 3 years and yielded -16.87\% return. This could be due to the stock ranking mechanism's poor performance in the period. In other words, for this period, the features we have selected were not effective indicators for stock performance. Additional features may be needed to improve stock ranking and selection performance. For example, we may consider implementing sentiment analysis of stocks. In addition, if this strategy were to be implemented, it may be meaningful to include a stop-loss mechanism. For example, when maximum drawdown reached 10\%, one may consider reevaluating features and select new features for the model.

\subsection{Feature Stability}

Another aspect to evaluate is weights assigned to each feature by the Lasso Regression at each period.

\begin{figure}[h!]
\centering
\includegraphics[width=0.8\textwidth]{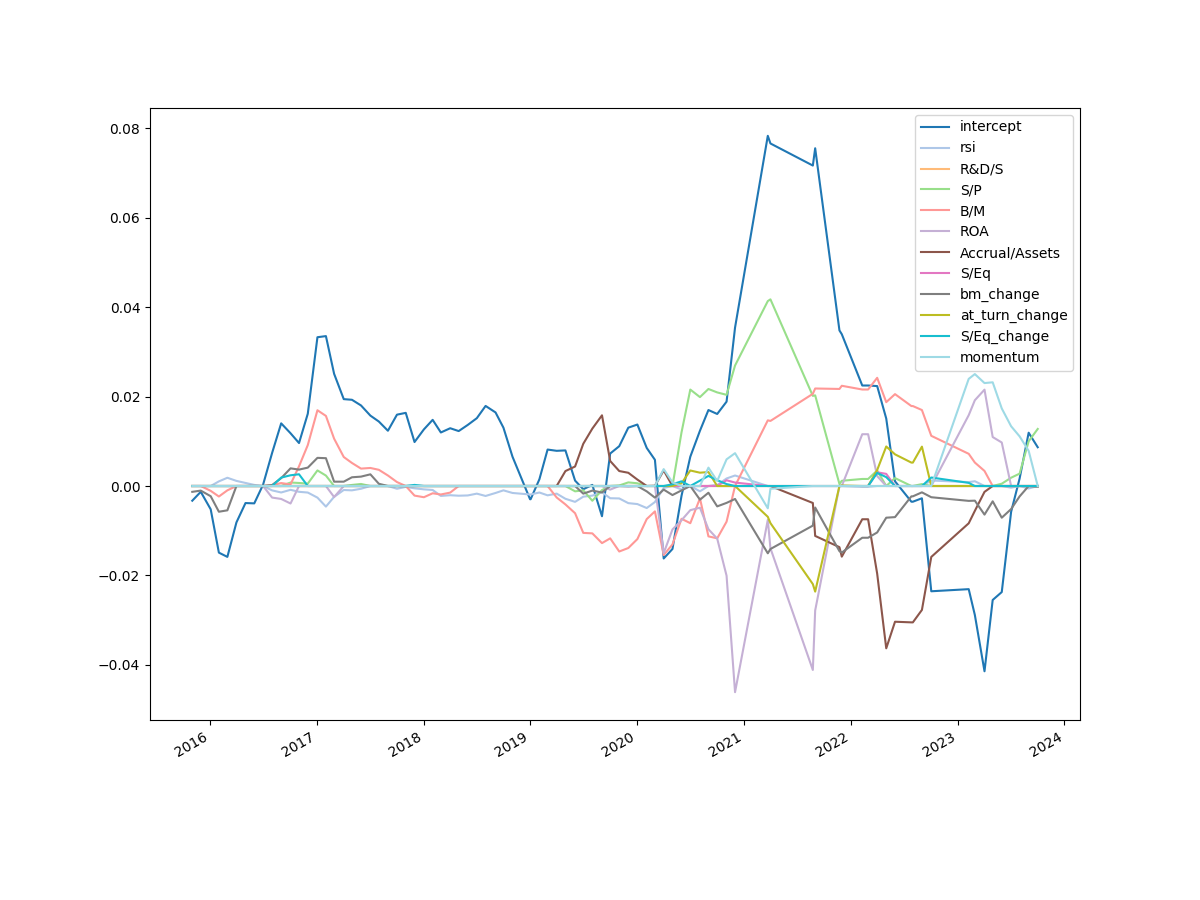}
\caption{Test Period Feature Weights.}
\label{fig:testweights}
\end{figure}

As seen in Figure \ref{fig:testweights}, the features appeared less stable towards later years. For example, signs of the features flipped more frequently in later years. This could be because the relationship found within training set no longer hold true in later years. As our train set and test set is over 10 years apart, this effect is probable. If this strategy were to be implemented, more frequent inspection or update on selected features may be needed.

Another point to note is that the weight of a number of features are set to 0 by the Lasso Regression. This could be because some features carry the same or similar information. Therefore, the model only used one of the features, and set weight of the rest to 0. It may be meaningful to perform hedging on the features to create orthogonal inputs for the model.

%% file: Summary.tex
\section{Summary}

The goal of this paper was to design and test a multi-factor market-neutral strategy for NYSE equities. First, we demonstrated how feature engineering and selection was performed. Three portfolio construction methods were suggested and backtested: risk parity, minimum variance beta neutral and equally weighted. Portfolio performance in the in-sample validation set indicated that risk parity offers higher risk-adjusted returns and a lower maximum drawdown, while maintaining market neutrality. The out-of-sample performance reinforced our conviction that the suggested strategy can deliver steady, low-risk returns.

%% file: main.bib
@online{cusip,
  author = {CUSIP Global Services},
  title = {About CGS Identifiers},
  year = {2024},
  url = {https://www.cusip.com/identifiers.html?section=CUSIP},
  urldate = {2024-12-01},
}

@article{rsi,
  title={The relative strength index revisited},
  author={{\c{T}}{\u{a}}ran-Moro{\c{s}}an, Adrian},
  journal={African Journal of Business Management},
  volume={5},
  number={14},
  pages={5855--5862},
  year={2011}
}

@book{tsi,
  title={Momentum, direction, and divergence},
  author={Blau, William},
  volume={5},
  year={1995},
  publisher={John Wiley \& Sons}
}

@article{trendclarity,
  title={Trended Momentum},
  author={Cai, Charlie X and Keasey, Kevin and others},
  year={2024}
}

@article{blitz2019characteristics,
  title={The characteristics of factor investing},
  author={Blitz, David and Vidojevic, Milan},
  journal={Journal of Portfolio Management},
  volume={45},
  number={3},
  pages={69--86},
  year={2019},
  publisher={Pageant Media}
}

@article{fama1996multifactor,
  title={Multifactor portfolio efficiency and multifactor asset pricing},
  author={Fama, Eugene F},
  journal={Journal of financial and quantitative analysis},
  volume={31},
  number={4},
  pages={441--465},
  year={1996},
  publisher={Cambridge University Press}
}

@article{arshanapalli1998multifactor,
  title={Multifactor asset pricing analysis of international value investment strategies},
  author={Arshanapalli, Bala and Daniel Coggin, T and Doukas, John},
  journal={Journal of Portfolio Management},
  volume={24},
  number={4},
  pages={10--+},
  year={1998},
  publisher={INSTITUTIONAL INVESTOR INC 488 MADISON AVENUE, NEW YORK, NY 10022 USA}
}

@article{jeancharles2013rpp,
  title={A Fast Algorithm for Computing High-Dimensional Risk Parity Portfolios},
  author={Théophile Griveau-Billion, Jean-Charles Richard, Thierry Roncalli},
  journal={SSRN},
  volume={24},
  number={4},
  pages={9},
  year={2013},
  publisher={SSRN}
}
